\documentstyle [12pt]{article}
\begin{document}
\baselineskip 20.0 pt
\par
\mbox{}
\vskip -1.25in\par
\mbox{}
 \begin{flushright}
\makebox[1.5in][1]{UU-HEP-91/19}\\
\makebox[1.5in][1]{November 1991}
 \end{flushright}
 \vskip 0.25in

\begin{center}
{\bf Nonlinear $\hat{W}_{\infty}$ Current Algebra in the
$SL(2,R)/U(1)$ Coset Model}\\
\vspace{40 pt}
{Feng Yu and Yong-Shi Wu}\\
\vspace{20 pt}
{{\it Department of Physics, University of Utah}}\\
{{\it Salt Lake City, Utah 84112, U.S.A.}}\\
\vspace{40 pt}
{\bf Abstract}\\
\end{center}
\vspace{10 pt}

Previously we have established that the second Hamiltonian structure
of the KP hierarchy is a nonlinear deformation, called
$\hat{W}_{\infty}$, of the linear, centerless $W_{\infty}$ algebra. In
this letter we present a free-field realization for all generators of
$\hat{W}_{\infty}$ in terms of two scalars as well as an elegant
generating function for the $\hat{W}_{\infty}$ currents in the
classical conformal $SL(2,R)/U(1)$ coset model. After quantization, a
quantum deformation of $\hat{W}_{\infty}$ appears as the hidden
current algebra in this model. The $\hat{W}_{\infty}$ current algebra
results in an infinite set of commuting conserved charges, which might
give rise to $W$-hair for the 2d black hole arising in the
corresponding string theory at level $k=9/4$.

\vspace{50 pt}

\newpage

1. {\it Introduction}~~~~~~
Very recently $W$-algebras of the infinite type, which extend the
conformal Virasoro algebra and incorporate all higher spin ($s\geq2$
or even $s\geq 1$) currents, have attracted a lot of attention in the
search of symmetries in $c=1, d=2$ strings. Those whose complete structure
is presently known include $w_{\infty}$ [1],
$W_{\infty}$ [2], $W_{1+\infty}$ [3] and $\hat{W}_{\infty}$ [4]. They may be
considered as the large $N$ limits [1] of the $W_N$ algebras of the finite
type [5], or believed to contain the latter upon reduction. In recent
topological approach [6] or in the KdV approach [7] to $d<2$ strings,
the string equation is shown to satisfy a set of (non-isospectral)
$W_{N}$ constraints [8] and to be invariant under the isospectral
$W_{N}$ transformations. In these models $d\rightarrow 2$ if
$N\rightarrow \infty$, so it is naturally hoped that the $d=2$ string
theory should be described by the KP hierarchy [9], which is known to
contain all (generalized) KdV hierarchies, and should embrace a
$W$-symmetry of the infinitr type in a similar manner. While the
relevance of the KP hierarchy remains to be clarified, this or that
$W$-algebra of the infinite type has indeed emerged or been proposed
as symmetry of $d=2$ string theory in several recent works [10,11,12,13,14].

To study a $W$-algebra containing infinitely many currents, choosing a
convenient basis is important. We will use the so-called KP basis
[4], in which the $W$-currents $u_r(z)$ appear as the coefficients in
the KP pseudo-differential (Psd) operator (with $u_{-1}=0$):
\begin{eqnarray}
L = D+\sum^{\infty}_{r=0}u_{r}D^{-r-1},~~~~~~~D\equiv \partial/\partial{z}
\end{eqnarray}
Previously it has been established that the first Hamiltonian structure
[15] of the KP hierarchy is none other than $W_{1+\infty}$ [16,17], and
its second Hamiltonian structure [18] is a nonlinear deformation [4]
of $W_{\infty}$, which we called $\hat{W}_{\infty}$. It is known that
$W_{\infty}$ has a two-boson realization [19]. In this letter we
construct a free-field realization for all $\hat{W}_{\infty}$ currents
in terms of two scalars, and show an elegant generating function for
the $\hat{W}_{\infty}$ currents in the classical confromal
$SL(2,R)/U(1)$ coset model. After quantization the current algebra in
the model naturally becomes a quantum version of $\hat{W}_{\infty}$.
Our work supplies the proof for the proposal and speculation made by
Bakas and Kiritsis [20] in a recent preprint, in which they have
presented evidence for the existence of a nonlinear $W$-algebra of the
infinite type in the same coset model and speculated it is a quantum
version of our $\hat{W}_{\infty}$. Recently Witten [21] has
interpreted the target space-time in the critical string theory
arising from the coset model at level $k=9/4$ as a 2d black hole. The
infinite set of commuting conserved chagres resulting from the
$\hat{W}_{\infty}$ current algebra in the coset model might give rise
to the so-called $W$-hair for the 2d black hole, whose existence and
consequence have been argued in ref.[22].

\vspace{5 pt}
2. {\it Free Boson Realization of $\hat{W}_{\infty}$}~~~~~~
Consider the currents $j(z)=\phi'(z)$, $\bar{j}(z)=\bar{\phi}'(z)$ of
two free scalars, satisfying
\begin{eqnarray} {\{\bar{j}(z),
\bar{j}(z')\}} ={\{j(z), j(z')\}} = 0,~~ {\{j(z), \bar{j}(z')\}}=
\partial_{z}\delta(z-z').
\end{eqnarray}
In our realization of $\hat{W}_{\infty}$ we express the KP operator
(1) in terms of the chiral currents
\begin{eqnarray} L = D+\bar{j}\frac{1}{D-(\bar{j}+j)}j
\equiv D+\bar{j}D^{-1}j+\bar{j}D^{-1}(\bar{j}+j)D^{-1}j+ \cdots .
\end{eqnarray}
In this way the $\hat{W}_{\infty}$ currents $u_{r}$ are realized by
functions of $\bar{j}$, $j$ and their derivatives. Notice the first
two terms, $D+\bar{j}D^{-1}j$, realize the linear $W_{\infty}$, while
the remaining terms represent its nonlinear deformation. For example,
the first a few currents are
\begin{eqnarray} u_{0} &=& \bar{j}j,~~~~~
u_{1} = -\bar{j}j'+\bar{j}j^{2}+\bar{j}^{2}j, \nonumber\\
 u_{2} &=&\bar{j}j''-3\bar{j}jj'-2\bar{j}^{2}j'-\bar{j}\bar{j}'j
+\bar{j}j^{3}+2\bar{j}^{2}j^{2}+\bar{j}^{3}j, \nonumber\\
u_{3} &=& -\bar{j}j'''+4\bar{j}jj''+3\bar{j}j'^{2}+3\bar{j}^{2}j''
+3\bar{j}\bar{j}'j'+\bar{j}\bar{j}''j
-6\bar{j}j^{2}j'-9\bar{j}^{2}jj'\nonumber\\ & &
-3\bar{j}\bar{j}'j^{2}-3\bar{j}^{3}j'
-3\bar{j}^{2}\bar{j}'j
+\bar{j}j^{4}+3\bar{j}^{2}j^{3}+3\bar{j}^{3}j^{2}+\bar{j}^{4}j.
\end{eqnarray}

To prove the representation (3) amounts to showing that the Poisson
brackets among $u_{r}$'s, evaluated according to eq.(2), are identical
to those of $\hat{W}_{\infty}$ as the second KP Hamiltonian structure.
First, for two functionals $\oint fdz$ and $\oint g(z')dz'$ of
$u_{r}$'s and their derivatives, their Poisson bracket from (2)
is given by
\begin{eqnarray} {\{\oint fdz, \oint gdz' \}} = \oint (\frac{\delta
f}{\delta \bar{j}} (\frac{\delta g}{\delta j})' + \frac{\delta
f}{\delta j} (\frac{\delta g}{\delta \bar{j}})')dz
\end{eqnarray}
where $(\delta g/\delta j)'$ can be replaced by
${[D-(\bar{j}+j), \delta g/\delta j ]}$. Then substitute
\begin{eqnarray}
\frac{\delta f}{\delta \bar{j}} &=& Res(\frac{1}{D-(\bar{j}+j)}j
\frac{\delta f}{\delta L}(1+\bar{j}\frac{1}{D-(\bar{j}+j)})), \nonumber\\
\frac{\delta f}{\delta j} &=& Res((1+\frac{1}{D-(\bar{j}+j)}j)
\frac{\delta f}{\delta L}\bar{j}\frac{1}{D-(\bar{j}+j)}),
\end{eqnarray}
etc. into eq.(5), where $Res P$ means the coefficient of the
$D^{-1}$ term in the operator $P$, and $\delta f/\delta L \equiv
\sum^{\infty}_{r=0} D^{r} (\delta f/\delta u_{r})$, with $\delta
f/\delta u_{r}$ being the usual variational derivative. Using the
calculus of Psd operators [9], we have been able to show
\begin{eqnarray}
{\{\oint fdz, \oint gdz' \}}
&=& \oint Res[L\frac{\delta f}{\delta L}(L\frac{\delta g}{\delta L})_{+}
-\frac{\delta f}{\delta L}L(\frac{\delta g}{\delta L}L)_{+} \nonumber\\
& & ~~~~
+ L\frac{\delta f}{\delta L}((L-D)\frac{\delta g}{\delta L})_{+}
-\frac{\delta f}{\delta L}L(\frac{\delta g}{\delta L}(L-D))_{+} ]dz.
\end{eqnarray}
This is exactly the defining brackets of $\hat{W}_{\infty}$ [4], since
the first two terms are actually the Dickey form [18] of the second KP
Hamiltonian structure with $u_{-1}\neq 0$, and the last two terms
represent the modification resulting from imposing the usual choice
$u_{-1}=0$ which is of second class. (The details of the proof will be
published elsewhere [23].)

\vspace{5 pt}
3. {\it Classical $\hat{W}_{\infty}$ Currents in
$SL(2,R)/U(1)$ Model} ~~~~~  Now let us proceed to show that our
$\hat{W}_{\infty}$ appears, as a hidden current algebra, in the
classical conformal $SL(2,R)_{k}/U(1)$ coset model.  Recall the free
field prescription [24] of the classical $SL(2,R)_{k}$ current algebra
with three bosons

\begin{eqnarray} J_{\pm} =
\sqrt{\frac{k}{2}}e^{\pm\sqrt{\frac{2}{k}}\phi_{3}} (\phi_{1}' \mp
i\phi_{2}')e^{\pm\sqrt{\frac{2}{k}}\phi_{1}},
\end{eqnarray}
Gauging $U(1)$
or taking the coset $SL(2,R)_{k}/U(1)$ is equivalent to restricting
the $U(1)$ current $J_{3}=0$ or simply $\phi_3=0$. Thus we are left
with $J_{\pm}$, which are just the bosonized $SL(2,R)_{k}/U(1)$
parafermion currents [25] (without loss of generality we set
$k=1$)
\begin{eqnarray}
\psi_{+}=\bar{j}e^{\bar{\phi}+\phi}, ~~~~
\psi_{-}=je^{-\bar{\phi}-\phi}
\end{eqnarray}
with two bosons $\phi_{1}=(1/\sqrt{2})(\phi +\bar{\phi})$ and
$\phi_{2}=(1/\sqrt{2}i)(\phi -\bar{\phi})$.

To generate higher-spin currents from them, we propose to consider the
bi-local product $\psi_{+}(z)\psi_{-}(z')$ and the expansion of this
ordinary product in powers of $z-z'\equiv \epsilon$ {\it{to all
orders}}. Our second main result is that this classical (rather than
operator) product expansion generates all $\hat{W}_{\infty}$ currents:
\begin{eqnarray}
\bar{j}e^{\bar{\phi}+\phi}(z)je^{-\bar{\phi}-\phi}(z') =
\sum^{\infty}_{r=0}u_{r}(z)\frac{\epsilon^{r}}{r!}.
\end{eqnarray}
with the coefficients $u_{r}(z)$ exactly the $\hat{W}_{\infty}$
currents previously realized by eq.(3).

To prove this, it is convenient to introduce the following definition:
The generating function of a Psd operator
$P(z)=\sum^{\infty}_{r=0}p_{r}(z) D^{-r-1}$ is a bi-local function
$F(z,z')=\sum^{\infty}_{r=0}p_{r}(z)(z-z')^{r}/r!$ , denoted by $P(z)
\Longleftrightarrow F(z,z')$. Then eq.(10) can be restated as
\begin{eqnarray}
\bar{j}\frac{1}{D-(\bar{j}+j)}j(z) \Longleftrightarrow
\bar{j}e^{\bar{\phi}+\phi}(z)je^{-\bar{\phi}-\phi}(z').
\end{eqnarray}
In fact, the right side depends only on the chiral currents $\bar{j}$
and $j$:
\begin{eqnarray}
 \bar{j}e^{\bar{\phi}+\phi}(z)je^{-\bar{\phi}-\phi}(z')
= \sum^{\infty}_{n=0}\sum^{\infty}_{k=0}\frac{(-1)^{n}}{n!k!}\bar{j}j^{(n)}
\epsilon^{n}(\sum^{\infty}_{m=0}\frac{(-1)^{m}}{(m+1)!}(\bar{j}+j)^{(m)}
\epsilon^{m+1})^{k}
\end{eqnarray}
where $j^{(n)}\equiv (\partial_{z}^{n}j)$, etc. For the left hand side one has
\begin{eqnarray}
\bar{j}\frac{1}{D-(\bar{j}+j)}j(z) = \sum^{\infty}_{k=0}\bar{j}
(D^{-1}(\bar{j}+j))^{k}D^{-1}j.
\end{eqnarray}
with the $k$-th term evaluated to be
\begin{eqnarray}
& & \sum^{\infty}_{m_{1},m_{2},...,m_{k},n=0}
\left( \begin{array}{c}
-1\\m_{1}
\end{array} \right)
\left( \begin{array}{c}
-m_{1}-2\\m_{2}
\end{array} \right) \cdots
\left( \begin{array}{c}
-m_{1}-m_{2}-\cdots -m_{k}-k-1\\n
\end{array} \right) \nonumber\\
& & \bar{j}(\bar{j}+j)^{(m_{1})}(\bar{j}+j)^{(m_{2})}\cdots
(\bar{j}+j)^{(m_{k})}j^{(n)}
D^{-m_{1}-m_{2}-\cdots -m_{k}-n-k-1} \nonumber\\
&=& \sum^{\infty}_{m_{1},m_{2},...,m_{k},n=0}
(-1)^{m_{1}+m_{2}+\cdots +m_{k}+n}
\frac{(m_{1}+m_{2}+\cdots +m_{k}+n+k)!}{(m_{1}+1)!(m_{2}+1)!\cdots
(m_{k}+1)!n!k!} \nonumber\\
& & \bar{j}(\bar{j}+j)^{(m_{1})}(\bar{j}+j)^{(m_{2})}\cdots
(\bar{j}+j)^{(m_{k})}j^{(n)}D^{-m_{1}-m_{2}
-\cdots -m_{k}-n-k-1}.
\end{eqnarray}
Here we have applied the identity $\left( \begin{array}{c} -a\\b
\end{array} \right) = (-1)^{b} \left( \begin{array}{c} a+b-1\\b
\end{array} \right) $ and then totally symmetrized the indices
$m_{i}$. Therefore we have, as desired,
\begin{eqnarray}
\bar{j}(D^{-1}(\bar{j}+j))^{k}D^{-1}j
\Longleftrightarrow
\sum^{\infty}_{n=0}\frac{(-1)^{n}}{n!k!}\bar{j}j^{(n)}
\epsilon^{n}(\sum^{\infty}_{m=0}\frac{(-1)^{m}}{(m+1)!}(\bar{j}+j)^{(m)}
\epsilon^{m+1})^{k}.
\end{eqnarray}

The generating function (10) allows us to obtain the complete
structure of $\hat{W}_{\infty}$ in a compact form. In fact, one can
generate all Poisson brackets among $\hat{W}_{\infty}$ currents by
\begin{eqnarray}
{\{\bar{j}e^{\bar{\phi}+\phi}(z)je^{-\bar{\phi}-\phi}(z-\epsilon),~
\bar{j}e^{\bar{\phi}+\phi}(w)je^{-\bar{\phi}-\phi}(w-\sigma)\}}
= \sum^{\infty}_{r,s=0}{\{u_{r}(z), u_{s}(w)\}}\frac{\epsilon^{r}
\sigma^{s}}{r!s!}
\end{eqnarray}
yielding the explicit and closed expression of ${\{u_{r}(z),
u_{s}(w)\}}$ given in [4].

\vspace{5 pt}
4. {\it Quantum $\hat{W}_{\infty}$ Currents in $SL(2,R)/U(1)$ Model}
{}~~~
{}From above discussions, it is natural to expect that the quantization
of the model will lead to a quantum version of $\hat{W}_{\infty}$.
However, merely changing the ordinary product in the left hand side of
eq.(10) to the operator product expansion and replacing terms
nonlinear in $j$ and $\bar{j}$ in the right hand side by normal
ordering are not enough to guarantee the closure of the algebra; as we
have emphasized before [4], since $\hat{W}_{\infty}$ is nonlinear, the
Jacobi identities for its central extension become central-charge
dependent and require non-trivial deformation in the non-central part
of the brackets.  Fortunately, the quantization of the model gives us
the necessary clue: In addition, we need to include the quantum
corrections into the relevant currents. For the $SL(2,R)_k$ currents,
we have [24,25,26]
\begin{eqnarray}
J_{\pm} = \sqrt{\frac{k}{2}}e^{\pm\sqrt{\frac{2}{k}}\phi_{3}}
(\phi_{1}' \mp i\sqrt{1-\frac{2}{k}}\phi_{2}')
e^{\pm\sqrt{\frac{2}{k}}\phi_{1}},
\end{eqnarray}
Thus the appropriate quantum $SL(2,R)_{k}/U(1)$ parafermionic currents
should be
\begin{eqnarray} \psi_{+}(p,z) &=&
\frac{1}{2}[(1+\sqrt{1-2p})\bar{j}+(1-\sqrt{1-2p})j]
e^{\sqrt{p}(\bar{\phi}+\phi)}, \nonumber\\
\psi_{-}(p,z) &=&
\frac{1}{2}[(1-\sqrt{1-2p})\bar{j}+(1+\sqrt{1-2p})j]
e^{-\sqrt{p}(\bar{\phi}+\phi)}
\end{eqnarray}
with $p=k^{-1}$. Their OPE, as the quantum version of eq.(10),
reads
\begin{eqnarray}
\psi_{+}(p,z)\psi_{-}(p,z-\epsilon)~ = ~
\epsilon^{-2p}\{\epsilon^{-2}+\sum^{\infty}_{r=0}u_{r}(p,z)
\frac{\epsilon^{r}}{r!}\}
\end{eqnarray}
which naturally generates all the quantum $\hat{W}_{\infty}(p)$ generators
in the KP basis. We note that the expression in the bracket
corresponds to the quantum KP operator:
\begin{eqnarray}
\epsilon^{-2}+\sum^{\infty}_{r=0}u_{r}(p,z)\frac{\epsilon^{r}}{r!}
\Longleftrightarrow D+\sum^{\infty}_{r=0}u_{r}(p,z)D^{-r-1} = L(p,z)
\end{eqnarray}
where we have set $\epsilon^{-2}\Longleftrightarrow D$. The OPE's for
$u_{r}(p,z)u_{s}(p,w)$, which manifest the complete structure of the
quantum $\hat{W}_{\infty}(p)$ algebra, may then be extracted from the
OPE of four $SL(2,R)_{k}/U(1)$ parafermionic currents:
\begin{eqnarray} L(p,z)L(p,w)
\Longleftrightarrow
\epsilon^{4p}(\psi_{+}(p,z)\psi_{-}(p,z-\epsilon))
(\psi_{+}(p,w)\psi_{-}(p,w-\sigma)).
\end{eqnarray}
The closure of the OPE's associated with the enveloping algebra of the
$SL(2,R)_{k}$ currents (17) in the neutral sector ensures the closure
of the quantum $\hat{W}_{\infty}(p)$ currents.

Explicitly we have calculated the first a few $u_{r}(p)$ in terms of
bosonic currents and verified that they are the same as in ref.[20] up
to a change of basis. For example, the first two currents are given by
(the notation for normal ordering being suppressed)
\begin{eqnarray}
u_{0}(p) &=&
(1-2p)\bar{j}j-\frac{1}{2}\sqrt{1-2p}\sqrt{p}(\bar{j}'-j'),
\nonumber\\ u_{1}(p) &=& -\frac{1}{2}((1-p)\sqrt{1-2p}+1-2p)\bar{j}j'
+\frac{1}{2} ((1-p)\sqrt{1-2p}-1+2p)\bar{j}'j \nonumber\\
& & -\frac{1}{2}p\sqrt{1-2p}(\bar{j}\bar{j}'-jj')
+\frac{1}{12}(3\sqrt{1-2p}-1) \sqrt{p}\bar{j}'' -
\frac{1}{12}(3\sqrt{1-2p}+1)\sqrt{p}j'' \nonumber\\
& & +(1-\frac{3}{2}p)\sqrt{p}(\bar{j}j^{2}+\bar{j}^{2}j)
+\frac{1}{6}p\sqrt{p} (\bar{j}^{3}+j^{3}).
\end{eqnarray}
(The expression for $u_2(p)$ is too long to exhibit here.) Their
OPE's are
\begin{eqnarray}
& & u_{0}(p,z)u_{0}(p,z') \sim (1-2p)(
\frac{2u_{0}(p,z')}{(z-z')^{2}}+\frac{u_{0}'(p,z')}{(z-z')}
+\frac{(1-2p-3p^{2})}{(z-z')^{4}}), \nonumber\\
& & u_{0}(p,z)u_{1}(p,z') \sim (1-2p)(
\frac{3u_{1}(p,z')}{(z-z')^{2}}+\frac{u_{1}'(p,z')}{(z-z')}
-\frac{2u_{0}(p,z')}{(z-z')^{3}} -\frac{(2-4p-6p^{2})}{(z-z')^{5}}),
\nonumber\\
& & u_{1}(p,z)u_{1}(p,z') \sim (1-\frac{3p}{2})
(\frac{4u_{2}(p,z')}{(z-z')^{2}}+\frac{2u_{2}'(p,z')}{(z-z')})
+\frac{2pu_{0}^{2}(p,z')}{(z-z')^{2}}
+\frac{pu_{0}^{2}{}'(p,z')}{(z-z')} \nonumber\\
& & +(1-p)(\frac{2u_{1}'(p,z')}{(z-z')^{2}}
+\frac{u_{1}''(p,z')}{(z-z')})
+\frac{8(1-p)pu_{0}(p,z')}{(z-z')^{4}}
+\frac{4(1-p)pu_{0}'(p,z')}{(z-z')^{3}} \nonumber\\
& & +\frac{(8-6p)pu_{0}''(p,z')}{5(z-z')^{2}}
+\frac{(7-4p)pu_{0}'''(p,z')}{15(z-z')}
-\frac{(3-14p)(4-7p)(1+p)}{3(z-z')^{6}}.
\end{eqnarray}
Thus, the OPE (21) determines a quantum deformation of the second KP
Hamiltonian structure. The classical limit
is recovered [26] by first
rescaling $j\rightarrow j/\sqrt{p}$, $u_r\rightarrow u_r/p$ and $[~,~]
\rightarrow p \{~,~\}$, then taking $p\rightarrow 0$.

\vspace{5 pt}
5. {\it Conserved $W$-charges and Application to 2d Black Hole}~~~~~~ The
classical $\hat{W}_{\infty}$ currents $u_{r}$ in the Hamiltonian basis
are defined by
\begin{eqnarray}
\hat{W}_{r}=\frac{1}{r-1}Res~L^{r-1}.
\end{eqnarray}
Their charges $\oint\hat{W}_{r}(z)dz$ are known [18,4] to be in
involution:
\begin{eqnarray}
{\{\oint\hat{W}_{r}(z)dz, \oint\hat{W}_{s}(z')dz'\}} = 0.
\end{eqnarray}
In particular, the charge of $\hat{W}_2$ is nothing but the
Hamiltonian of the model. This implies the existence of infinitely
many conserved $W$-charges in the conformal $SL(2,R)/U(1)$ coset
model. The first a few classical charges are explicitly given in
ref.[4].

Upon quantization, the Poisson brackets in eq.(25) are replaced by the
commutators:
\begin{eqnarray}
{[\oint\hat{W}_{r}(p,z)dz, \oint\hat{W}_{s}(p,z')dz']} = 0
\end{eqnarray}
with appropriate quantum modifications in the relation (24)
between the $\hat{W}_{\infty}(p)$ currents in the Hamiltonian and the KP
bases. We have constructed the first a few quantum $\hat{W}_{\infty}(p)$
currents whose charges commute with each other:
\begin{eqnarray}
\hat{W}_{2}(p) &=& \frac{1}{(1-2p)}u_{0}(p), \nonumber\\
\hat{W}_{3}(p) &=& u_{1}(p)+\frac{1}{2}u_{0}'(p), \nonumber\\
\hat{W}_{4}(p) &=& u_{2}(p)+u_{1}'(p)+\frac{(5+4p)}{15}u_{0}''(p)
+\frac{p}{(1-2p)}u_{0}^{2}(p),
\end{eqnarray}
consistent with ref.[20]. In particular, $\hat{W}_{2}(p)$ is the
energy-momentum tensor in the quantized coset model. It would be
interesting to obtain a compact expression for the quantum version of
eq.(24) for all $r$.

At $k=9/4$, the conformal $SL(2,R)_k/U(1)$ coset model [27] has
central charge $c=26$ and, therefore, gives rise to a critical $d=2$
string theory, whose space-time interpretation is a 2d black hole
[21]. The above-derived infinite set of commuting $W$-charges implies
certain infinite target-space symmetries in both the classical and
quantum coset model. As usual in string theory, one may expect that
these symmetries on the world-sheet can be elevated to stringy gauge
symmetries in target space-time associated with higher-spin states.
If so, the quantum mechanical states of the black hole may be
classified by their conserved charges (the so-called $W$-hair). And if
the classification is complete the existence of $W$-hair should, as
have been argued in ref.[22], ensure the maintenance of quantum
coherence even during black-hole evaporation.

Finally, we note that the conserved charges we have identified are
those in the sigma-model theory; the relation of them with the
$W$-symmetries [28, 10-14] formulated in the $c=1$ matrix model
remains to be clarified.  On the other hand, we feel the emergence of
the KP structure in the present treatment strongly suggests the
possibility of a fundamental (perhaps quantum-deformed) KP approach
to, at least $c=1$, non-perturbative string theory.

\vspace{5 pt}
{\it{Acknowledgements}}:~ The authors thank I. Bakas and C. J. Zhu for
discussions. The work is supported in part by NSF grant PHY-9008452.
F. Y. is also supported in part by Utah Graduate Research Fellowship.

\vspace{40 pt}
\begin{center}
{\large REFERENCES}
\end{center}
\begin{itemize}
\vspace{5 pt}

\item[1.] I. Bakas, Phys. Lett. 228B (1989) 57; Comm. Math. Phys. 134
(1990) 487.
\item[2.] C. Pope, L. Romans and X. Shen, Phys. Lett. 236B (1990) 173;
Nucl. Phys. B339 (1990) 191.
\item[3.] C. Pope, L. Romans and X. Shen, Phys. Lett. 242B (1990) 401.
\item[4.] F. Yu and Y.-S. Wu, Utah preprint UU-HEP-91/09, May 1991;
to appear in Nucl. Phys. B.
\item[5.] A. B. Zamolodchikov, Theor. Math. Phys. 65 (1985) 1205.
\item[6.] E. Witten, Nucl. Phys. B340 (1990) 281;
R. Dijkgraaf and E. Witten, Nucl. Phys. B342 (1990) 486.
\item[7.] M. Douglas, Phys. Lett. 238B (1990) 17;  T. Banks, M. Douglas,
N. Seiberg and S. Shenker, Phys. Lett. 238B (1990) 279;
D. Gross and A. Migdal, Nucl. Phys. B340 (1990) 333;
P. Di Francesco and D. Kutasov, Nucl. Phys. B342 (1990) 589.
\item[8.] E. Verlinde and H. Verlinde, Nucl. Phys. B348 (1991) 457;
R. Dijkgraaf, E. Verlinde and H. Verlinde, Nucl. Phys. B348 (1991) 435;
M. Fukuma, H. Kawai and R. Nakayama, Int. J. Mod. Phys. A6 (1991)
1385; Tokyo/KEK preprint UT-572; J. Goeree, Nucl. Phys. B358 (1991) 737.
\item[9.] M. Sato, RIMS Kokyuroku 439 (1981) 30; E. Date, M. Jimbo, M.
Kashiwara and T. Miwa, in Proc. of RIMS Symposium on Nonlinear Integrable
Systems, eds. M. Jimbo and T. Miwa, (World Scientific, Singapore, 1983);
G. Segal and G. Wilson, Publ. IHES 61 (1985) 1.
\item[10.] E. Witten, IAS preprint IASSNS-HEP-91/51.
\item[11.] A. M. Polyakov and I. Klebanov, Princeton preprint.
\item[12.] G. Moore and N. Seiberg, Rutgers/Yale preprint RU-91-29 and
YCTP-P19-91.
\item[13.] J. Avan and A. Jevicki, Brown preprints BROWN-HET-801,824 (1991).
\item[14.] S. R. Das, A. Dhar, G. Mandal and S. R. Wadia, ETH/IAS/Tata
preprint ETH-TH-91/30, IASSNS-HEP-91/52 and TIFR-TH-91/44.
\item[15.] Y. Watanabe, Ann. di Mat. Pura Appl. 86 (1984) 77.
\item[16.] F. Yu and Y.-S. Wu, Phys. Lett. 263B (1991) 220.
\item[17.] K. Yamagishi, Phys. Lett. 259B (1991) 436.
\item[18.] L. A. Dickey, Annals New York Academy of Sciences, (1987) 131.
See also J. Figueroa-O'Farill, J. Mas and E. Ramos, Leuven preprint
KUL-TF-91/23, May 1991.
\item[19.] I. Bakas and E. Kiritsis, Nucl. Phys. B343 (1990) 185.
\item[20.] I. Bakas and E. Kiritsis, Maryland/Berkeley/LBL preprint
UCB-PTH-91/44, LBL-31213 and UMD-PP-92/37, Sept. 1991.
\item[21.] E. Witten, Phys. Rev. D44 (1991) 314.
\item[22.] J. Ellis, N. Mavromatos and D. Nanopoulos, CERN preprints
CERN-TH-6147/91, Jun. 1991; CERN-TH-6229/91, Sept. 1991.
\item[23.] F. Yu and Y.-S. Wu, Utah preprint, to appear.
\item[24.] A. Gerasimov, A. Marshakov and A. Morozov, Nucl. Phys. B328
(1989) 664.
\item[25.] O. Hern\'andez, Phys. Lett. 233B (1989) 355.
\item[26.] To recover the classical limit (8) one first sets
$k\rightarrow k\hbar^{-1}$, rescales the whole currents by a
$\hbar^{1/2}$ factor, and then lets $\hbar\rightarrow 0$. Also recall
that each field $\phi_{i}$ implicitly has the dimension $\hbar^{1/2}$.
\item[27.] L. Dixon, J. Lykken and M. Peskin, Nucl. Phys. B235 (1989)
215.
\item[28.] D.J. Gross and I.R. Klebanov, Nucl. Phys. B352 (1991) 671.

\end{itemize}

\end{document}